

\magnification=1200
\hsize=13cm
\vskip 2cm
\centerline{\bf 1. Introduction}
\vskip 0.7cm

The discovery of high temperature superconductivity [1]
has determined an increasing interest in the study of
strongly correlated electron systems, and in particular of the
Hubbard model [2] and the $t$-$J$ model [3].
The Hubbard model describes the dynamics of non-relativistic
electrons moving on a lattice. Its Hamiltonian
consists of a kinetic hopping term of strength $t$, and
an on-site repulsion between up and down spins of strength $U$
modeling the Coulomb interaction.
At infinite $U$, one must pay infinite energy to
put two electrons of opposite spins in the same point. Hence,
double occupancy is strictly forbidden, and
the electrons can move in the lattice only if
some sites are vacant. Consequently, at half-filling, {\it i.e.} when the
number of electrons equals the number of lattice points, the infinite
$U$ Hubbard model
describes an insulator (of the Mott type) [4].
Notice that this insulating behavior occurs at half-filling,
while ordinary insulators are always characterized by completely filled
bands.

The Hubbard model at half-filling is known to describe
a Mott insulator also when $U$ is large but finite,
{\it i.e.} in the strong-coupling limit [4].
In this case one can develop (degenerate)
perturbation theory in the small parameter $t^2/U$
and still use states with no
double electron occupancy. To leading
order the Hubbard interaction transforms into an
antiferromagnetic Heisenberg interaction with a coupling constant
$J \propto t^2/U$. Thus, to this order
the Hubbard system can be
effectively described by a Hamiltonian with a kinetic
hopping of strength $t$ and an antiferromagnetic Heisenberg spin-exchange
interaction of strength $J$, both acting on states without
double electron occupancy. This is the so-called $t$-$J$ model.
Of course, at half-filling the $t$-$J$ model reduces to the
Heisenberg model since the hopping term
becomes ineffective, and describes a Mott insulator
with an antiferromagnetic N\'eel order. If the band
filling factor is lowered below 1/2 by the
introduction of holes ({\it i.e.} by doping), the system
becomes a conductor, and eventually the N\'eel order
gives way to superconductivity. Thus we can say that
the $t$-$J$ model below half-filling
represents a conducting doped antiferromagnet
which may become a superconductor [3].

In one space dimension, both the Hubbard model and the $t$-$J$
model (with $J=\pm2\,t$ or with $J/t\to 0$)
can be exactly solved by the Bethe {\it Ansatz}
[5,6,7], and the spectrum of their collective excitations
can be completely determined.
In particular for the $t$-$J$ model below half-filling,
two kinds of quasi-particle gapless
excitations are identified above the anti-ferromagnetic ground
state: one carries only charge and no spin and is called holon,
the other carries only spin and no charge and is called spinon.
In other words there is a separation between the spin and charge
degrees of freedom. Even though this phenomenon can be
rigorously established only using the exact solution provided by the
Bethe {\it Ansatz}, it is believed that
the spin-charge separation
actually occurs also in non-exactly solvable models (like for example the
one-dimensional $t$-$J$ model with $J\not=\pm2\,t$) and even
in higher dimensions [8,9]. Therefore, in
order to understand better the mechanism of the spin-charge
separation, it would be desirable to
develop techniques and formalisms that capture the essential dynamical
features of strongly correlated electron systems
without insisting on their exact solvability.

One way of implementing the spin-charge separation  is to
use the formalism of the slave operators [10] and write
the electrons as products of one spinless antiholon and one
neutral spinon of spin 1/2. This formalism is very useful since
it allows the application to electron systems below half-filling
of several standard techniques that would be otherwise
inapplicable, such as path-integrals,
mean field approximations, large $N$ expansions and so on. The
primary reason for this is that the anholonomic constraint
which characterizes an electron system
below half-filling, is transformed into a holonomic
constraint on the slave operators (see for instance
(2.1) and (2.9) below). Another fundamental
property of this formalism is that from the outset
it introduces the local freedom of choosing arbitrarily
the phase of holons and spinons which therefore
are naturally related to abelian gauge theories. On the basis
of this very general observation, one expects that the effective
action describing the low-energy dynamics of holons and spinons
should be invariant under local gauge
transformations. This is indeed what happens
in many cases, like for example in the $t$-$J$ model
at half-filling, {\it i.e.} in the Heisenberg model [11,12].

In this paper we show that this happens also
in the $t$-$J$ model below half-filling, when both spin and
charge degrees of freedom can be excited.
In particular, by using the slave operators
in the so-called ${\rm CP}^1$ representation [13,14]
we prove that
the Hamiltonian of the one-dimensional $t$-$J$ model
in the continuum limit gives rise to an Euclidean field theory
in two dimensions that is explicitly invariant under
abelian gauge transformations. More precisely, this
theory is a ${\rm CP}^1$ model with topological term [15,16,17]
for the spin degrees of freedom minimally coupled to
a massless Dirac fermion describing the charge excitations.

Our results are similar to those of previous investigations
on the dynamics of doped antiferromagnets in
one or two dimensions [18,19,20,21]. However, a more detailed comparison
with the existing literature (and specifically
with [18]) shows several significant differences.
Firstly, we use from the beginning the slave operator
formalism in the ${\rm CP}^1$ representation, and deduce everything
from the Hamiltonian of the $t$-$J$ model in such representation.
Thus, the coupling between holons and spinons is not
imposed by hand, but it is derived directly from the hopping
term of the Hamiltonian of the $t$-$J$ model.
Secondly, we introduce only one type of holons (like it happens
in the exact solution of the Bethe {\it Ansatz}), and do not have
the need of distinguishing between two types of holons with
opposite charges as in [18,19,20]. This is an explicit consequence
of using the ${\rm CP}^1$ representation, as we will see in detail.
Finally, in our end result we find a four-fermion interaction for the
holon field that previously was not emphasized. However, the general
structure of our approach as well as some motivations are in
agreement with what already exists in the literature.

This paper is organized as follows: In section 2 we introduce
the ${\rm CP}^1$ representation for the $t$-$J$ model
in any dimension and write the Hamiltonian explicitly
in terms of holons and spinons. In Section 3
we study the low-energy effective action for the
charge and spin degrees of freedom in one dimension, and show
that in the continuum limit this action becomes
that of a ${\rm CP}^1$ model with a topological
term minimally coupled to a massless Dirac
fermion. The bosonic field of the ${\rm CP}^1$ model
describes the dynamics of the spin waves produced
by the spinons, while the fermionic field represents
the low-energy holon excitations.
Finally in Section 4 we rewrite the effective action
by introducing an abelian gauge field to exhibit explicitly
the local gauge invariance of the model, and then present our
conclusions.

\vskip 2cm
\centerline{\bf 2. The ${\rm CP}^1$ Representation of the $t$-$J$ Model}
\vskip 0.7cm
The cuprate oxides which some years ago allowed
the discovery of high temperature
superconductivity [1], share a common structural
feature: the presence of
layers made of copper and oxigen ions where all electric
transport phenomena
(and eventually superconductivity) seem to take place [8].
Each of these Cu-O layers can be represented
as a lattice with the copper ions sitting on the sites and
the oxigen ions on the bonds.
In undoped materials every copper ion has one unpaired electron
in its outer $3d$ shell, whilst all electrons in the outer $2p$
shell of the oxigen ions are paired. Therefore  the Cu-O layer
of an undoped oxide can be regarded as a lattice with
exactly one electron per site, {\it i.e.} at half-filling.
Clearly, in this case no electric current can flow and
the system is an insulator.
Spectroscopic experiments reveal that the spins of the
unpaired electrons are antiferromagnetically ordered
at low temperatures [22], and only spin-exchange interactions can
occur. Thus, the physics of these undoped materials can be effectively
described
by an antiferromagnetic Heisenberg model [8].

Upon doping, some oxigen ions of the Cu-O layers lose one
electron from their outer $2p$ shells so that new unpaired
spins show up. However the hybridization between the
$2p$ oxigen orbitals and the $3d$ copper orbitals strongly
binds each one of these unpaired O electrons to one of
the unpaired Cu electrons already present, in such a way
that a spin singlet is formed [3]. Therefore the Cu-O layer of
a doped oxide can be represented as a lattice where most
of the sites are occupied by one electron and the remaining
few sites are occupied by a (charged) spin singlet. The
unpaired O electrons introduced by
doping can actually move in the lattice
and couple to different Cu electrons. Thus an electric current
is generated and the material becomes a conductor
and eventually a superconductor below a
high critical temperature.

The electron dynamics in these doped
cuprate oxides can
be effectively described by the so-called $t$-$J$ model which
was originally proposed by Zhang and Rice [3]; together
with the Hubbard model [2], it has become one of the most
studied examples of strongly correlated electron systems.
Let us denote by ${\hat c}_\alpha^\dagger(i)$ and ${\hat c}_\alpha(i)$
the fermionic operators which create and destroy an
electron at site $i$ with $z$-component of the spin $\alpha/2$
($\alpha=\pm1$). Since each site can accomodate
at most one electron, we must require that
$$
n(i) \equiv \sum_{\alpha=\pm1} {\hat c}_\alpha^\dagger(i)
{}~{\hat c}_\alpha(i) \leq 1 \ \ ,
\eqno(2.1)
$$
{\it i.e.} we must exclude double occupancy of
any lattice point. The spin-1/2
operators which generate local rotations on the electron spins,
can be represented
\`a la Schwinger by
$$
{\hat S}^a(i) = {1\over 2}\sum_{\alpha,\beta=\pm1}
{\hat c}_\alpha^\dagger(i)\,
\sigma^a_{\alpha\beta}\,
{\hat c}_\beta(i)
{}~~~~,~~~~ a=1,2,3\ \ ,
\eqno(2.2)
$$
($\sigma^a$ being the Pauli matrices). It can be easily
checked that the ${\hat S}^a(i)$'s
satisfy the
standard $SU(2)$ algebra.
In terms of these operators, the Hamiltonian of the
$t$-$J$ model with chemical
potential $\mu$ can be written as
$$
\eqalign{
H_{tJ} =~& {\cal P} \left[
-t\sum_{<i,j>}\sum_{\alpha=\pm1}
\Big({\hat c}_\alpha^\dagger(i)\,{\hat c}_\alpha(j)+
{\rm h.\,c.}\Big)
\right. \cr
&~~~~\left.+J\sum_{<i,j>}\left({\bf {\hat S}}(i)
\cdot{\bf {\hat S}}(j)
-{1\over 4}n(i)\,n(j)\right)+\mu
\sum_{i}n(i)\right]
{\cal P}\ \ , }
\eqno(2.3)
$$
where ${\cal P}$ is the Gutzwiller projection operator [23]
enforcing the constraint (2.1) and the symbol ${<i,j>}$ denotes
a pair of nearest-neighbor lattice sites.
The parameters $t$ and $J$ are coupling
constants which we take to be positive: $t$ is
the strength of the kinetic hopping term, and
$J$ is the strength of the spin-exchange
antiferromagnetic interaction. The chemical potential $\mu$ in (2.3)
must be fixed in such a way that
$$
\big\langle n(i) \big\rangle = 1 - \delta\ \ ,
\eqno(2.4)
$$
where the doping concentration $\delta$ is defined by
$$
\delta = {{N - N_{\rm el}}\over {N}}
\eqno(2.5)
$$
with $N_{\rm el}$ representing the number of unpaired electrons
and $N$ the number of lattice sites. At half-filling when $\delta=0$,
the $t$ term
of (2.3) becomes ineffective
and only the $J$ term survives,
so that in the absence of doping the $t$-$J$ model reduces to the
(antiferromagnetic) Heisenberg model. The situation is clearly
different below half-filling, where also the $t$ term gives
a non-trivial contribution
to the electron dynamics.
However, it should be emphasized that because
of the Gutzwiller projection,
the $t$ and $J$ terms of the Hamiltonian (2.3)
always act disjointedly. In fact, given a
couple of nearest-neighbor
points, when both of them are occupied by one electron only
the $J$ term acts;
on the other hand, when one of the two sites is empty the $J$
term vanishes
while the hopping term $t$ is effective; finally when both sites
are empty the $t$ and $J$ terms are both vanishing.

Several authors suggested that at small doping ({\it i.e.}
$\delta \simeq 0.1 - 0.2$) a very
interesting phenomenon should
occur: the spin and charge degrees of freedom should separate, indicating that
the whole system behaves as a Luttinger liquid [24].
In other words, the spectrum of the lowest lying excitations of the $t$-$J$
model should consist of holons,
which carry only charge and no spin, and spinons, which carry only
spin and no charge [8,9].
Actually, a rigorous proof of the occurrence of this spin-charge
separation exists
only for the one-dimensional $t$-$J$ model since this can be exactly solved
by Bethe Ansatz, at least when $J=\pm2\,t$ (supersymmetric point) [6]
or when $J/t \to 0$ [7]. For arbitrary values of $t$ and $J$ or for
the two-dimensional model, there are no exact
results available, but some approximate solutions [25] as well as
recent numerical calculations [26] seem to support the conjecture
that the spin-charge separation always occurs, and also that the
peculiar Luttinger liquid properties of the one-dimensional model survive
in higher dimensions.

One common way to formally implement the spin-charge separation
is to use slave operators [10] and factorize the electron oscillators,
which obviously carry both spin and charge, into products of
holons and spinons. Specifically one writes
$$
{\hat c}_\alpha(i) = {\hat s}_\alpha(i)\,
{\hat h}^\dagger(i)~~~~,~~~~
{\hat c}^\dagger_\alpha(i) = {\hat h}(i)\,
{\hat s}^\dagger_\alpha(i)\ \ ,
\eqno(2.6)
$$
where ${\hat h}^\dagger(i)$ and ${\hat h}(i)$ are
respectively the creation and annihilation operators
for charged spinless holons, and ${\hat s}^\dagger_\alpha(i)$
and ${\hat s}_\alpha(i)$ are the creation and
annihilation operators
for neutral spinons of spin $\alpha/2$.

The statistics of holons and spinons is not {\it a priori}
determined. However, since the electron oscillators must be fermionic, there
are in general only two possibilities: one
is to take fermionic holons and bosonic spinons,
the other is to take bosonic
holons and fermionic spinons. In either case, the products of
one holon and one spinon
(like those in (2.6)) are always fermionic. Actually, in two dimensions
where the statistics may be anything [27], it can be shown that holons and
spinons may be anyons of arbitrary complementary statistics [28].
In this paper we choose to work with fermionic
holons and bosonic spinons that
obey the following (anti)commutation relations
$$
\left\{{\hat h}(i)\,,\,{\hat h}^\dagger(j)\right\} = \delta(i,j)
{}~~~~~,~~~~~
\left[{\hat s}_\alpha(i)\,,\,{\hat s}^\dagger_\beta(j)\right] =
\delta_{\alpha\beta}~\delta(i,j)\ \ ,
\eqno(2.7)
$$
where $\delta(i,j)$ is the lattice $\delta$-function.
The reason for this choice is that, with fermionic
holons,
$$
{\hat c}_\alpha(i)\,{\hat c}_\beta(i) =
{\hat c}^\dagger_\alpha(i)\,{\hat c}^\dagger_\beta(i)=0
\eqno(2.8)
$$
for {\it all} $\alpha$ and $\beta$. Thus, double occupancy is always
automatically forbidden at any lattice site as required by (2.1).

Another important point to emphasize is that holons and spinons are {\it not}
completely independent but must be constrained if one wants to recover
the correct electronic configurations. One possibility to achieve this
goal is to impose that
$$
{\hat h}^\dagger(i)\,{\hat h}(i)+
{\hat s}^\dagger_\alpha(i)\,{\hat s}_\alpha(i) = 1\ \ .
\eqno(2.9)
$$
(From now on,
the summation symbol over repeated spin indices will be understood.)
The holon and spinon operators subject to this relation form the so-called
slave fermion representation \footnote{$^1$}{When (2.9) is imposed on bosonic
holons and fermionic spinons one realizes the so-called slave boson
representation.}.
Notice that (2.9) implies that each lattice site is always occupied
either by a holon or by a spinon of spin up or down, and consequently
 at each point
there are only three
possible states, just as required by (2.1).
Thus, in this slave operator formalism the anholonomic
constraint (2.1) is transformed into
the equality constraint (2.9). This is certainly a remarkable improvement.

In this representation the spin operators
${\hat S}^a(i)$  can be written only in
terms of spinon oscillators as follows (see {\it e.g.} [28])
$$
{\hat S}^a(i) = {1\over 2}
{\hat s}_\alpha^\dagger(i)
\,\sigma^a_{\alpha\beta}\,
{\hat s}_\beta(i)
{}~~~~,~~~~ a=1,2,3\ \ .
\eqno(2.10)
$$
Upon using (2.6) and (2.10), and taking into account the constraint
(2.9), the Hamiltonian of the $t$-$J$ model in the slave fermion representation
is
$$
H_{tJ} = H_{t}+H_{J}+\mu\,\sum_{i}
\Big(1-{\hat h}(i)^\dagger\,
{\hat h}(i)\Big)\ \ ,
\eqno(2.11)
$$
where
$$
H_{t} = -t\,\sum_{<i,j>}
\Big({\hat h}(i)\,{\hat s}_\alpha^\dagger(i)\,{\hat s}_\alpha(j)
\,{\hat h}^\dagger(j)
+{\rm h.\,c.}\Big)\ \ ,
\eqno(2.12)
$$
and
$$
\eqalign{
H_{J}
=&{J\over 4}\,\sum_{<i,j>}\Bigg[\sum_{a=1}^3\Big(
{\hat s}_\alpha^\dagger(i)\,
\sigma^a_{\alpha\beta}\,{\hat s}_\beta(i)\,
{\hat s}_{\alpha '}^\dagger(j)\,
\sigma^a_{\alpha ' \beta '}\,{\hat s}_{\beta '}(j)\Big)
\cr
&~~~~~~~~~~~~~~~~~~~~~-{\hat s}_\alpha^\dagger(i)\,
{\hat s}_\alpha(i)\,
{\hat s}_{\alpha '}^\dagger(j)\,
{\hat s}_{\alpha '}(j)\Bigg]\ \ .}
\eqno(2.13)
$$
The theory described by (2.11), (2.12) and (2.13)
can be studied in the mean
field approximation; however the reliability of this analysis
is questionable and the results obtained with this method are
purely qualitative since it is very hard to treat the {\it local}
constraint (2.9) in a systematic and controllable way.

One possibility to overcome this problem is to
partially liberate the slave holons and spinons, and use the so-called
${\rm CP}^1$ representation [13,14] in which only the spinons
are constrained while the holons are left free. More precisely,
the ${\rm CP}^1$ representation is characterized by the constraint
$$
{\hat s}^\dagger_\alpha(i)\,{\hat s}_\alpha(i)=1\ \ ,
\eqno(2.14)
$$
meaning that each lattice site is {\it always} occupied by a spinon;
on the contrary, the holons may or may not be present. The average
of the holon occupation number is directly related to the doping
concentration $\delta$; in fact
$$
\big\langle n(i) \big\rangle = \big\langle h(i)
\,h^\dagger(i)\,s^\dagger_\alpha(i)\,s_\alpha(i)\big\rangle=
1 -\big\langle h^\dagger(i)h(i)\big\rangle \ \ ,
\eqno(2.15)
$$
and by comparison with (2.4), we immediately deduce that
$$
\big\langle h^\dagger(i)h(i)\big\rangle = \delta\ \ .
\eqno(2.16)
$$
We want to stress that there is a crucial
difference between the slave operator representation considered
earlier and
the ${\rm CP}^1$ representation introduced now. Indeed, as noted above,
in the slave operator description the Hilbert space of holons
and spinons at each lattice point is three-dimensional like the original
electron
Hilbert space; instead,
in the ${\rm CP}^1$ representation holons and spinons
form a four-dimensional space, because each point must always have one
spinon (up
or down) but may or may not have a holon.
To remove this
discrepancy, one should correct
the ${\rm CP}^1$ representation with a suitable projection operator $P$
in such a way that the Hilbert space of
holons and spinons is reduced from four to three dimensions. Then, instead of
(2.6) one should write [29]
$$
{\hat c}_\alpha(i) = P\,{\hat s}_\alpha(i)\,
{\hat h}^\dagger(i)\,P^\dagger~~~~,~~~~
{\hat c}^\dagger_\alpha(i) =  P\,{\hat h}(i)\,
{\hat s}^\dagger_\alpha(i)\,P^\dagger
\eqno(2.17)
$$
with the constraint (2.14). A
systematic treatment of
this operator $P$ is cumbersome and difficult.
However, it has been recently shown [29] that even without
the projection
one can obtain very good results both from a qualitative and
a quantitative point of view.
This is the same attitude that we take here, since
we will not insert $P$ in any formulas. It is
precisely the freedom we gain in
this way that allows us to make further progress.

In the ${\rm CP}^1$ representation the spin
operators ${\hat S}^a(i)$'s  are
realized with both holon and spinon oscillators as follows
$$
{\hat S}^a(i) = {1\over 2}
\Big( 1-{\hat h}^\dagger(i)\,{\hat h}(i)\Big)\,
{\hat s}_\alpha^\dagger(i)
\,\sigma^a_{\alpha\beta}\,
{\hat s}_\beta(i)
{}~~~~,~~~~ a=1,2,3\ \ .
\eqno(2.18)
$$
When no holon is present, ${\hat S}^a(i)$ receives a
contribution only by the spinons, just like
in (2.10);
on the contrary, due to the prefactor, the spin vanishes
in all sites occupied
by holons \footnote{$^2$}{Despite the appearance,
the same is true
for the slave fermion representation (2.10) if the constraint
(2.9) is taken into account.}.  Then,
whenever ${\hat h}^\dagger(i)\,{\hat h}(i)=1$,
the spinon configuration
can be arbitrarily chosen without changing the physical
spin ${\hat S}^a(i)$ which remains zero; in other words, when the holons
are present the ${\rm CP}^1$ spinons are merely fictitious
degrees of freedom which
can be used at ease. This fact will play a crucial role in
the following.

With this in mind, the Hamiltonian of the $t$-$J$ model in the
${\rm CP}^1$ representation can be written as in (2.11) with
$H_t$ given by (2.12) and $H_J$ by
$$
\eqalign{
H_{J}
={J\over 4}\,&\sum_{<i,j>}\Bigg\{
\Big( 1-{\hat h}^\dagger(i)\,{\hat h}(i)\Big)\,
\Big( 1-{\hat h}^\dagger(j)\,{\hat h}(j)\Big)
\cr
&~~~~~~~~\cdot\left[\sum_{a=1}^3\Big(
{\hat s}_\alpha^\dagger(i)\,
\sigma^a_{\alpha\beta}\,{\hat s}_\beta(i)\,
{\hat s}_{\alpha '}^\dagger(j)\,
\sigma^a_{\alpha ' \beta '}\,{\hat s}_{\beta '}(j)\Big)-1\right]\Bigg\}\ \ .}
\eqno(2.19)
$$
and with the spinons subject to the relation (2.14).
To implement systematically
this local constraint one may add to the Hamiltonian the following
term
$$
\sum_i \lambda(i)\,\Big({\hat s}^\dagger_\alpha(i)\,{\hat s}_\alpha(i)-1\Big)
\ \ ,
\eqno(2.20)
$$
where $\lambda(i)$ serves as a Lagrange multiplier
over which we will integrate. However, before this integration
is done the spinons are
not constrained, and thus we can use the coherent state quantization method
to analize the theory. The coherent states for
holons and spinons, $|h\rangle$ and $|s\rangle$, are defined in such a way
that
$$
\eqalign{
{\hat h}(i)\,|h\rangle &= h(i)\,|h\rangle~~~~,~~~~
\langle h|\,{\hat h}^\dagger(i) =
\langle h|\,h^*(i)\ \ , \cr
{\hat s}_\alpha(i)\,|s\rangle &= s_\alpha(i)\,|s\rangle~~~~,~~~~
\langle s|\,{\hat s}_\alpha^\dagger(i) =
\langle s|\,s_\alpha^*(i)\ \ , }
\eqno(2.21)
$$
where $h(i)$ and $s_\alpha(i)$ are respectively
anticommuting and commuting complex fields.

Then, the partition function
of the $d$-dimensional $t$-$J$ model at temperature
$1/\beta$ becomes an Euclidean path integral in $d+1$
dimensions
$$
{\cal Z}_{tJ} = \int\!{\cal D}^2\!h~
{\cal D}^2\!s_1~{\cal D}^2\!s_2~{\cal D}\lambda
{}~~{\rm e}^{-S_{tJ}}\ \ ,
\eqno(2.22)
$$
where
$$
\eqalign{
S_{tJ}= \int_0^\beta \!d\tau
&\left[\sum_i\Big( h^*(i,\tau)\,\partial_{\tau}h(i,\tau) -
s^*_\alpha(i,\tau)\,\partial_{\tau}s_\alpha(i,\tau)\Big) +
{\cal H}_{tJ}\right.
\cr
&\left.~+\sum_i\,{\lambda}(i,\tau)\Big(
s^*_\alpha(i,\tau)\,s_\alpha(i,\tau) - 1\Big)
\right]\ \ .}
\eqno(2.23)
$$
In this formula $\tau$ is an imaginary ``time''
parameter on which the fields
$h$ and $s$ are made dependent,
${\cal H}_{tJ}$ is the functional
obtained from $H_{tJ}$ in the ${\rm CP}^1$ representation
by replacing every occurrence of ${\hat h}(i)$,
${\hat h}^\dagger(i)$, ${\hat s}_\alpha(i)$ and
${\hat s}^\dagger_\alpha(i)$ with
$h(i,\tau)$, $h^*(i,\tau)$, $s_\alpha(i,\tau)$ and
$s^*_\alpha(i,\tau)$ respectively according to
(2.21), and finally ${\lambda}(i,\tau)$ is the Lagrange multiplier which is
integrated
along the imaginary axis from $-{\rm i}\infty$ to
$+{\rm i}\infty$ to enforce the local constraint
$$
s^*_\alpha(i,\tau)\,s_\alpha(i,\tau) = 1 \ \ .
\eqno(2.24)
$$
This is nothing but the expectation value
in a coherent state of the original operator constraint
(2.14), and also explains
why (2.17) is called
the ${\rm CP}^1$ representation.

We remark that the $\tau$-derivative terms in the first line
of (2.23) represent the Berry phase for the holon and spinon
fields which is usual in the coherent state quantization
method \footnote{$^3$}{The minus sign in the spinon term is the same that
appears in [30].}. Notice however that
the spinon phase has a physical meaning
only in those points that are not occupied by holons,
{\it i.e.} where there
is a physical spin to which the spinon is directly related.
In the other points where holons are present, the spinon phase
has no direct significance since it corresponds to a fictitious
spinon configuration. Finally, we recall
that one has to impose periodic boundary
conditions in $\tau$ on the bosonic fields, and antiperiodic
boundary conditions on the fermionic ones in order to
incorporate correctly the statistics effects.

At any site $i$, the spinon field $s_\alpha(i)$ can be used
to define a three-dimensional unit vector ${\bf n}(i)$ with components
$$
n^a(i) =
s_\alpha^*(i)
\,\sigma^a_{\alpha\beta}\,
s_\beta(i)
{}~~~~,~~~~ a=1,2,3\ \ .
\eqno(2.25)
$$
If no holon is present at $i$, then ${\bf n}(i)$ points along
the spin direction
(cf (2.18)). Thus, an antiferromagnetic arrangement of
the spins corresponds to an antiferromagnetic order
for the vector field ${\bf n}(i)$; we call this a N\'eel
configuration. We now assume that the lattice where
the model is defined is bipartite so that no frustration is
present \footnote{$^4$}{This choice is customary in the
study of antiferromagnetism and moreover the lattice
representing the Cu-O layers of the cuprate oxides is
in this class.}. Then, we can distinguish between even
and odd sites, which we denote by $a$ and $b$
respectively.
In a classical N\'eel configuration we have
$$
{\bf n}(b) = - {\bf n}(a)\ \ .
\eqno(2.26)
$$
This can be realized in terms of the spinon field by choosing
for instance
$$
s_\alpha(b) = \varepsilon_{\alpha\beta}\,s^*_\beta(a)\,
{\rm e}^{{\rm i}\theta}\ \ ,
\eqno(2.27)
$$
where $\varepsilon_{\alpha\beta}$ is
the completely antisymmetric tensor of $SU(2)$ and $\theta$
is an arbitrary phase. We notice that (2.27)
corresponds to the usual choice of placing
a certain spin representation on the even sites
and its conjugate one on the odd sites [30].

Let us assume for a moment that (2.27) holds
in the whole lattice, even in those
points where holons are present and the spin is vanishing.
If this is the case, then
the holons are frozen and cannot move.
This is most clearly seen by reinstating temporarily
the holon operators so that the
hopping Hamiltonian $H_t$ becomes
$$
H_{t}=-t\sum_{<a,b>}\Big(
{\hat h}(a)\,s^*_\alpha(a)\,s_\alpha(b)\,
{\hat h}^\dagger(b)+{\rm h. c.} \Big)\ \ .
\eqno(2.28)
$$
If the spinons are arranged according to (2.27), then the
hopping matrix elements are obviously vanishing. Therefore, one can say
that the holons cannot move in a rigid N\'eel background and
must distort the neighboring spinons in order to
acquire kinetic energy. This is a rather well known fact [31,14].
We point out that in
the gauge theory of
[18,19,20] the hopping term (2.28) is substituted by hand with
a next-to-nearest
neighbor hopping Hamiltonian which forces the holons
to jump only within one
sublattice (even or odd) where the spinons are
ferromagnetically ordered.
However, in such a case one loses contact with
the original $t$-$J$ model
in which a nearest-neighbor hopping term is the only one
that is present to give kinetic energy to the holons.

This puzzle can be overcome if one recalls that at the
holon sites the spinon configuration is {\it a priori}
not well defined. Therefore, assuming that (2.27)
holds everywhere on the lattice is a too strong
statement that is not really justified. In fact,
an antiferromagnetic spinon arrangement is natural
only for those links between two occupied sites,
but not for those bonds connecting (at least) one
holon. For example, let us consider the electron
configuration represented in Fig. 1a. Using the
holon-spinon language in the ${\rm CP}^1$ representation
we can describe Fig. 1a by saying that the even site $a$
is occupied by a spinon of spin up and no holon, while
the odd site $b$ is occupied by a holon {\it and} a
spinon. It seems that in principle there is no
preferred choice for the orientation of the latter spinon since the
spin is vanishing at $b$. However, if we consider
the situation after the hole has hopped (Fig. 1b)
we see that an up-spin appears in $b$ and thus the
spinon in $b$ should be chosen upwards ({\it i.e.} in
the same direction of the spinon in $a$).
With this in mind, it is then clear that
the spinons must be {\it chosen} in a ferromagnetic order on every
link involving one holon and one spinon but must be in an antiferromagnetic
configuration on those links without holons.
However, for ease of notation and later convenience, it
would be better not to distinguish between different
kinds of links and, if possible, to deal only with one uniform order
(either ferromagnetic or antiferromagnetic).

We can achieve this goal by flipping the physical spins on all odd sites.
(This is customary in the study of antiferromagnetism.)
After this is done, it is obvious that the spinon configuration
becomes ferromagnetic in all occupied sites, and then it is no
problem to choose the spinons on the empty sites as before
in such a way that the spinon configuration be uniformly
ferromagnetic.
This procedure corresponds to making the following change of variable
$$
s_\alpha(b) \longrightarrow \varepsilon_{\alpha\beta}\,s_\beta^*(b)\Big(
1-{\hat h}^\dagger(b)\,{\hat h}(b)\Big) + s_\alpha(b)\,
{\hat h}^\dagger(b)\,{\hat h}(b)
\eqno(2.29)
$$
in the Hamiltonian. From the explicit expressions, one can easily see that
the $t$ term of $H_{tJ}$ does not change under (2.29) while the $J$ term
picks up an overall minus sign.

Also the spinon Berry phase is modified by (2.29). Indeed, we have
$$
\eqalign{
-s^*_\alpha(b,\tau)\partial_{\tau}s_\alpha(b,\tau)\longrightarrow
&-s_\alpha(b,\tau)\partial_{\tau}s^*_\alpha(b,\tau)\Big(
1-{\hat h}^\dagger(b)\,{\hat h}(b)\Big)\cr
&-s^*_\alpha(b,\tau)\partial_{\tau}s_\alpha(b,\tau)\,
{\hat h}^\dagger(b)\,{\hat h}(b)\ \ .}
\eqno(2.30)
$$
{}From the constraint (2.24) it follows that
$s_\alpha(b,\tau)\partial_{\tau}s^*_\alpha(b,\tau) =
-\,s^*_\alpha(b,\tau)\partial_{\tau}s_\alpha(b,\tau)$, and
thus the contribution of the
odd sites to the spinon Berry phase can be written as
$$
\sum_b
\Big( 1-2h^*(b,\tau)\,h(b,\tau)\Big)\,
s^*_\alpha(b,\tau)\partial_{\tau}s_\alpha(b,\tau)\ \ .
\eqno(2.31)
$$
Since the even sites are not affected
by the spin flips, their contribution to the spinon phase remains as before.

Putting everything together, writing
explicitly the sums over even and odd sites whenever is necessary, restoring
the functional notation also for the holons and
dropping the $\tau$-dependence of the fields for ease of
notation, we finally arrive
at the following action
$$
\eqalign{
S_{tJ} =&\int_0^\beta \!\!d\tau
\Bigg\{\!-\sum_a s^*_\alpha(a)\partial_{\tau}s_\alpha(a)
+\sum_b
\Big( 1-2h^*(b)\,h(b)\Big)\,
s^*_\alpha(b)\partial_{\tau}s_\alpha(b)
\cr
&~~~~~+\sum_i \Big[h^*(i)\partial_{\tau}h(i)
\,+\,\lambda(i)\Big(
s^*_\alpha(i)\,s_\alpha(i) - 1\Big)\Big]
+{\cal H}_{tJ}
\Bigg\}\ \ ,}
\eqno(2.32)
$$
where
$$
{\cal H}_{tJ} = {\cal H}_{t}+{\cal H}_{J}+\mu\sum_i\Big(1-
h^*(i)\,h(i) \Big)\ \ ,
\eqno(2.33)
$$
with
$$
{\cal H}_{t}=-t\sum_{<a,b>}\Big(
h(a)\,s^*_\alpha(a)\,s_\alpha(b)\,h^*(b)
+{\rm c. c.} \Big)\ \ ,
\eqno(2.34)
$$
and
$$
{\cal H}_J = -{J\over 2}\sum_{<a,b>}\Big(1-
h^*(a)\,h(a)\Big)\Big(1-
h^*(b)\,h(b)\Big)~s^*_\alpha(a)\,s_\alpha(b)\,s^*_\beta(b)
\,s_\beta(a)\ \ .
\eqno(2.35)
$$
Notice the overall minus sign in ${\cal H}_J$ that, as noted above,
originates directly from
the spin flip on the odd sites of the lattice realized by (2.29).
Furthermore, in order to obtain
the explicit expression (2.35) we used standard identities on the
Pauli matrices.
Finally, we remark again that in the action (2.32) the spinon
fields are classically ordered in a ferromagnetic way
everywhere.

This is the ${\rm CP}^1$ representation of the $t$-$J$ model.
When no holes are
present ({\it i.e.} $h^*(i)\,h(i)=0$ everywhere), (2.32) reduces to the
well known ${\rm CP}^1$ representation of the Heisenberg model [32].
In the next section we will study the action (2.32) in the particular
case of a one-dimensional lattice.

\vskip 2cm
\centerline{\bf 3. The Continuum Field Theory Description}
\centerline{\bf of the $t$-$J$ Model in One Dimension}
\vskip 0.7cm
The action (2.32) is a field theory description for the
$t$-$J$ model that is valid
in any dimension. However, to make further progress,
from now on we will restrict our considerations only to a
one-dimensional lattice ({\it
i.e.} a chain) where some exact results can be obtained.

Let us begin our analysis by first setting $J=0$ in
(2.32) \footnote{$^5$}{We recall that
when $J=0$ the $t$-$J$ model is equivalent to a
Hubbard model with infinite on site repulsion.}.
If we reinstate temporarily the holon operators,
the Hamiltonian becomes
$$
H_{tJ}^{J=0} =-t\sum_{<a,b>}\Big(
{\hat h}(a)\,s^*_\alpha(a)\,s_\alpha(b)\,
{\hat h}^\dagger(b) +{\rm h. c.} \Big) -\mu\sum_i
{\hat h}^\dagger(i)\,{\hat h}(i) +\mu\,N \ \ ,
\eqno(3.1)
$$
where $N$ is the number of points in the chain.
(Hereinafter we will drop the additive constant $\mu\,N$
of (3.1).)
We can interpret $H_{tJ}^{J=0}$ as the
Hamiltonian for the hopping of holons
in a background given by the spinons, which,
as explained in the
previous section,
must be ferromagnetically ordered. If the spinons were constant
classical fields,
this would mean that
$$
s_\alpha(a) = s_\alpha(b)
\eqno(3.2)
$$
for any pair of nearest neighbor points $a$ and $b$, and thus,
because of (2.24) the Hamiltonian
(3.1) would reduce simply to
$$
{H'}_{tJ}^{J=0}=-t\sum_{<a,b>}\Big(
{\hat h}(a)\,{\hat h}^\dagger(b)
 +{\rm h. c.} \Big) -\mu\sum_i
{\hat h}^\dagger(i)\,{\hat h}(i) \ \ .
\eqno(3.3)
$$
This is the Hamiltonian of a fermionic tight-binding model which
can be easily diagonalized by Fourier transform. After this is
done, one can see that it describes free
fermions with dispersion relation
$$
\epsilon(k) = 2\,t\,\cos \,(k\,\ell) -\mu\ \ ,
\eqno(3.4)
$$
where $\ell$ is the lattice spacing
and the momentum $k$ lies in the first
Brillouin zone between 0 and $2\,\pi$.
If one requires that
$$
\big\langle h^\dagger(i)h(i)\big\rangle = \delta
\eqno(3.5)
$$
(cf (2.16)), the chemical potential
at zero temperature must be fixed as
$$
\mu= -2\,t\,\cos \,(\pi\,\delta)\ \ .
\eqno(3.6)
$$
Correspondingly, in the first Brillouin zone there
exist two Fermi points given by
$$
k_{\rm F}^\pm = {1\over \ell}~\pi\,(1\pm\delta)
\eqno(3.7)
$$
around which the dispersion relation $\epsilon(k)$
is linear with a Fermi velocity
$$
v_{\rm F}^\pm=-2\,t\,\ell\,\sin k_{\rm F}^\pm =
\pm 2\,t\,\ell\,\sin (\pi\,\delta)\ \ .
\eqno(3.8)
$$
These are the typical values for spinless fermions with concentration
$\delta$, and moreover they are in full agreement with the
exact Bethe Ansatz values of the one-dimensional
infinite-$U$ Hubbard model to
which the $t$-$J$ model reduces at $J=0$ .

In the theory described by (3.3) it is rather straightforward
to compute the correlation function
$\big\langle h(a)\,h^\dagger(b)\big\rangle$
for any pair of nearest neighbor points $a$ and $b$ under
the assumption that it does not depend on the link
$(a,b)$ and is real. Indeed, by computing the derivative
with respect to $t$ of the free energy associated
to ${H'}_{tJ}^{J=0}$ and taking into account
(3.6), one finds that at zero temperature
$$
\chi \equiv \big\langle h(a)\,h^\dagger(b)\big\rangle
= {\sin (\pi\,\delta) \over \pi}\ \ .
\eqno(3.9)
$$
We remark that this same result could be obtained in the
more general case in which the spinons are not constant by
using the mean field approximation.

Since in physical processes one can excite only the particles whose momentum
is close to the Fermi surface, we linearize our
theory near the Fermi points
(3.7) (for definiteness we choose $k_{\rm F}^+$ which
from now on we denote
simply by $k_{\rm F}$), and write the holon operator as
$$
{\hat h}(i) = {\rm e}^{+{\rm i}k_{\rm F} i}\,{\hat\psi}_+(i)
+{\rm e}^{-{\rm i}k_{\rm F} i}\,{\hat\psi}_-(i)\ \ ,
\eqno(3.10)
$$
where ${\hat\psi}_\pm(i)$  are smoothly varying
on the lattice since the Fermi factors
${\rm e}^{\pm{\rm i}k_{\rm F} i}$
have been pulled out. Then, upon inserting (3.10) into (3.3), and
using (3.6) and (3.8), we get
$$
\eqalign{
{H'}_{tJ}^{J=0}=&~{\rm i}\,{v_{\rm F}\over \ell}
\sum_b\Big[{\hat\psi}^\dagger_-(b)\,{\hat\psi}_-(b+\ell)
-{\hat\psi}^\dagger_+(b)\,{\hat\psi}_+(b+\ell)\cr
&~~~~~~~~~~-{\hat\psi}^\dagger_-(b)\,{\hat\psi}_-(b-\ell)+
\,{\hat\psi}^\dagger_+(b)\,{\hat\psi}_+(b-\ell)\Big]\ \ .}
\eqno(3.11)
$$
Notice that in deriving (3.11) we have dropped
all terms with any rapidly
oscillating factor according to the standard
procedure [4], and also that
the sum over all odd sites actually reproduces
the sum over the
whole lattice.

The Euclidean action that is equivalent to (3.11)
in the functional formalism
is, in obvious notations,
$$
\eqalign{
{S'}_{tJ}^{J=0}=&
\int_0^\beta \!d\tau ~\Bigg\{
\sum_i \Big[\psi^*_-(i,\tau)\partial_{\tau}\psi_-(i,\tau)+
\psi^*_+(i,\tau)\partial_{\tau}\psi_+(i,\tau)\Big]\cr
&~~~~~+{\rm i}\,{v_{\rm F}\over \ell}
\sum_b\Big[\psi^*_-(b,\tau)\,\psi_-(b+\ell,\tau)
-\psi^*_+(b,\tau)\,\psi_+(b+\ell,\tau)\cr
&~~~~~~~~~~~~-\psi^*_-(b,\tau)\,\psi_-(b-\ell,\tau)+
\,\psi^*_+(b,\tau)\,\psi_+(b-\ell,\tau)\Big]\Bigg\}\ \ .}
\eqno(3.12)
$$
It is now possible to take the continuum limit of (3.12). In order
to do this correctly, we must remember
that in one space dimension a fermionic field has
engineering dimension
$1/2$. Therefore, we first replace $\psi_\pm$ with
$\sqrt{\ell}\,\psi_\pm$ so that
$$
{1\over \ell}\,\psi^*_+(b,\tau)\,\psi_+(b\pm\ell,\tau) \longrightarrow
\psi^*_+(b,\tau)\,\psi_+(b,\tau)
\pm\ell \,\psi^*_+(b,\tau)\partial_1\psi_+(b,\tau)\ \ .
\eqno(3.13)
$$
(A similar relation obviously holds for $\psi_-$).
Then, for notational convenience we write $\partial_\tau$ as
$\partial_0$, and in the limit $\ell\to 0$ we replace
$\left(2\,\ell\int\limits_0^\beta\!
d\tau\sum\limits_b\right)$ with $\int \!d^2x$.
As a result we find
$$
{S'}_{tJ}^{J=0}=\int \!d^2x~\Big\{{\bar \psi}(x)\gamma^0\partial_0
\psi(x) + v_{\rm F}\,{\bar \psi}(x)\gamma^1\partial_1
\psi(x)\Big\}\ \ ,
\eqno(3.14)
$$
where we have introduced the notation
$$
\psi(x)\equiv\left(\matrix{
 \psi_-(x) \cr
 \psi_+(x) \cr}\right)
{}~~~,~~~{\bar \psi}(x)\equiv\psi^\dagger(x)\,
\gamma^0=\Big(\psi_+^*(x)~,~\psi_-^*(x)\Big)
\eqno(3.15)
$$
with $\gamma^0=\sigma^1$ and $\gamma^1=\sigma^2$ being $\,2\times2\,$ Euclidean
gamma matrices. Eq. (3.14) represents the action of
a ``relativistic'' massless Dirac fermion in two dimensions with a
characteristic velocity $v_{\rm F}$.
We have given the details of
this standard derivation [4] so that it will be simpler
to follow the later developements.

So far the spinons have been considered as a classical constant
background in which
the holons move like free fermions. If we want to take into
account the presence of the spinons at the quantum level, we must study
the Hamiltonian (3.1) with the spinons yielding a
configuration that is not
constant as in (3.2) but can fluctuate. Thus, the
spinons should become true dynamical degrees of
freedom which can take all possible values
compatible with the constraint (2.24). However, in a
sort of semiclassical approximation, we assume
that the spinons despite their fluctuations, still
define a ferromagnetic order in the lattice, which
is broken only by (possibly antiferromagnetic)
deviations that
are small and of the order of the lattice
spacing $\ell$.
Thus, following [11], for any site $i$ we posit
$$
s_\alpha(i) = z_\alpha(i)\,\sqrt{1-\ell^2\,
\Delta_\beta^*(i)\,\Delta_\beta(i)} + \ell\,\Delta_\alpha(i)\ \ ,
\eqno(3.16)
$$
where $z_\alpha$ is a complex slowly varying
field such that
$$
z^*_\alpha(i)\,z_\alpha(i) = 1\ \ ,
\eqno(3.17)
$$
and
$$
z_\alpha(i\pm\ell) \simeq z_\alpha(i)
\pm \ell\,\partial_1 z_\alpha(i)\ \ ,
\eqno(3.18)
$$
and $\Delta_\alpha$ is a small staggered fluctuation
such that
$$
z^*_\alpha(i)\,\Delta_\alpha(i)+\Delta^*_\alpha(i)\,z_\alpha(i)=0 \ \ ,
\eqno(3.19)
$$
and
$$
\Delta_\alpha(i\pm \ell) \simeq - \Delta_\alpha(i) \ \ .
\eqno(3.20)
$$
Notice that the square root factor in (3.16) together with
(3.17) and (3.19), preserves
the normalization of $s_\alpha$ to all orders in $\ell$.

The meaning of these equations is quite straightforward.
In fact, from (3.16) and (3.18) one can see that to order $\ell^0$
that there is a ferromagnetic alignment among the spinons
because $s_\alpha(a)\simeq s_\alpha(b)$. However
to order $\ell^1$ the spinon alignment is deformed both by the gradient
of $z_\alpha$ and by the
small fluctuation $\Delta_\alpha$
which explicitly introduces an antiferromagnetic
short range distorsion due to the staggering
sign of (3.20). Of course we must remember that in those
sites that are not occupied by holons, the ferromagnetic
order of the $z_\alpha$'s actually corresponds to a physical
antiferromagnetic spin configuration, while the antiferromagnetic
behavior of field $\Delta_\alpha$ actually corresponds to a physical
ferromagnetic deviation from a local N\'eel state.
This is because we flipped the spinons on the odd sublattice
(see (2.29)).

We now linearize the theory near the Fermi level of the holons
and insert the decomposition (3.10) into the Hamiltonian (3.1)
(from now on we will work only in the functional formalism, so that
obvious change of notation should be made in these formulas). A simple
calculation
leads to the following result
$$
\eqalign{
{\cal H}_{tJ}^{J=0} =&~
t\sum_b\Bigg\{\Big[{\rm e}^{+{\rm i}k_{\rm F}\ell}
\,\psi_+^*(b-\ell)\,\psi_+(b)\,s^*_\alpha(b)\,s_\alpha(b-\ell)\cr
&~~~~~~~~~~+
{\rm e}^{-{\rm i}k_{\rm F}\ell}
\,\psi_+^*(b+\ell)\,\psi_+(b)\,s^*_\alpha(b)\,s_\alpha(b+\ell)
+{\rm c.c.}\Big]\cr
&~~~~~~~~~~+\big(\psi_+\leftrightarrow\psi_-~~,~~k_{\rm F}
\leftrightarrow -k_{\rm F}\big)\Bigg\}\cr
&\!\!\!-{\mu\over 2} \sum_b\Bigg\{
\Big[\psi_+^*(b-\ell)\,\psi_+(b-\ell)+
\psi_+^*(b+\ell)\,\psi_+(b+\ell)\cr
&~~~~~~~~~~+2\,\psi_+^*(b)\,\psi_+(b)\Big]
+ \big(\psi_+\leftrightarrow\psi_-\big)\Bigg\} \ \ .}
\eqno(3.21)
$$
Then, we decompose the spinon fields
according to (3.16) and perform
the continuum limit along the
lines we discussed before, by keeping
at most terms of order $\ell^2$. Notice that
a bosonic field in one space dimension has zero engineering dimensions
and thus no redefinition
of the spinons is necessary. On the contrary, as noted above,
the fermionic holons must be rescaled by a factor
of $\sqrt{\ell}$ and therefore the square root factors
of (3.16) can be simply put to one since they would
yield contributions of higher order in $\ell$.
Taking this into account and using (3.18) and
(3.20), it is possible to
show that inside the curly brackets of (3.21)
all the terms involving the fluctuation
field $\Delta_\alpha$
exactly cancel and so do also the terms of order $\ell$.
Then,
the final result can be written in the continuum
limit as
$$
\int _0^\beta \!d\tau~{\cal H}_{tJ}^{J=0} = v_{\rm F} \int d^2x~
\Big\{
{\bar \psi}(x)\gamma^1\big[\partial_1 +{\rm i}\,a_1(x)\big]
\psi(x)\Big\}\ \ ,
\eqno(3.22)
$$
where we have used the notation (3.15) and defined
$$
a_1(x) = {\rm i} \,z^*_\alpha(x)\,\partial_1z_\alpha(x)\ \ .
\eqno(3.23)
$$
Because of the constraint (3.17), $a_1(x)$ is real and looks like
the space component of a gauge field to which the holons are
minimally coupled. As we will see later this is indeed
the correct interpretation.

Let us now consider the terms of the action $S^{J=0}_{tJ}$ which involve
$\tau$ derivatives. From (2.32) we see that these can be distinguished
into a spinon Berry phase
$$
S_{\rm B} = \int_0^\beta \!d\tau~\Bigg\{\sum_b
s^*_\alpha(b)\,\partial_\tau s_\alpha(b)
-\sum_a s^*_\alpha(a)\,\partial_\tau s_\alpha(a)
\Bigg\}\ \ ,
\eqno(3.24)
$$
which does not involve holons, and into the remainder
$$
\eqalign{
S_{\rm h} =& \int_0^\beta \!d\tau\,\sum_b \Bigg\{ {1\over 2}\Big[
h^*(b-\ell)\,\partial_\tau h(b-\ell)
+h^*(b+\ell)\,\partial_\tau h(b+\ell)\Big]
+h^*(b)\,\partial_\tau h(b) \cr
&~~~~~~~~~~~~~~~
-2h^*(b)\,h(b)\,s^*_\alpha(b)\,
\partial_\tau s_\alpha(b) \Bigg\}\ \ ,
}
\eqno(3.25)
$$
which instead contains the holon field.
The continuum limit of $S_{\rm h}$ can be
easily performed as before
and one gets
$$
S_{\rm h} = \int\!d^2x~\Big\{
{\bar \psi}(x)\gamma^0\big[\partial_0 +{\rm i}\,a_0(x)\big]
\psi(x)\Big\}\ \ ,
\eqno(3.26)
$$
where
$$
a_0(x) = {\rm i} \,z^*_\alpha(x)\,\partial_0z_\alpha(x)\ \ .
\eqno(3.27)
$$
As expected, (3.26) is the natural completion of (3.22).
Thus we can say
that in the continuum limit the action
of the one-dimensional $t$-$J$ model with $J=0$ is given by [33]
$$
S_{tJ}^{J=0}= {\tilde S}_{\rm B} + S_\psi =
{\tilde S}_{\rm B}+\int\!d^2x~\Big\{
{\bar \psi}\big(\,{\slash\!\!\!\partial}
+{\rm i}\,{\slash \!\!\!a}\big)
\psi\Big\}\ \ ,
\eqno(3.28)
$$
where ${\tilde S}_{\rm B}$ is the continuum
limit of the spinon Berry
phase (3.24) to which we will return in a moment,
and the ``~/~'' notation means
$$
\slash\!\!\!\partial=\gamma^0\,\partial_0+
v_{\rm F}\,\gamma^1\,\partial_1
{}~~~,~~~
\slash\!\!\!a=\gamma^0\,a_0+
v_{\rm F}\,\gamma^1\,a_1\ \ .
\eqno(3.29)
$$
We remark that $S_\psi$ is
invariant under the following (standard)
gauge transformation
$$
\eqalign{
\psi(x) &\longrightarrow \psi'(x)={\rm e}^{-{\rm i}
\Lambda(x)}~\psi(x)\cr
a_\mu(x)&\longrightarrow a'_\mu(x)=a_\mu(x)+\partial_\mu
\Lambda(x)~~~~~,~~\mu=0,1}
\eqno(3.30)
$$
for an arbitrary function $\Lambda(x)$.
Of course, in order to interpret (3.30) as a true
local symmetry of the full action
$S_{tJ}^{J=0}$ we need to examine in detail also the spinon Berry phase.
Actually, the properties of ${\tilde S}_{\rm B}$ have already
been investigated at length by many authors
following [11]. Thus, we simply recall the main
results referring to the original
literature for their derivation [11,12,30,34].
If we substitute (3.16) into (3.24) and use (3.18) and (3.20),
we can easily obtain
$$
{\tilde S}_{\rm B} = S_{\rm top} + S'\ \ ,
$$
where
$$
S'= \int\!d^2x~\Big\{
\Delta_\alpha^*(x)\,\partial_0z_\alpha(x) -
\partial_0z_\alpha^*(x)\,\Delta_\alpha(x)\Big\}\ \ ,
\eqno(3.31)
$$
and $S_{\rm top}$ is the continuum limit of
$$
\int_0^\beta \!d\tau~\Bigg\{
\sum_b z^*_\alpha(b)\,\partial_\tau z_\alpha(b)
-\sum_a z^*_\alpha(a)\,\partial_\tau z_\alpha(a)
\Bigg\}\ \ .
\eqno(3.32)
$$
At first sight it would seem that this expression
vanishes; indeed since the $z_\alpha$'s are ferromagnetically
ordered, the contributions of the two sub-lattices
seem to cancel each other. However, this cancellation occurs only
for the bulk of the lattice leaving possible residual contributions
at the boundaries. In fact,
a careful analysis of (3.32) in the continuum
limit (see for example [18,30]
for details) leads to
$$
S_{\rm top} = {{\rm i}\over 2}\int\!d^2x ~
\Big\{\varepsilon^{\mu\nu}\,\partial_\mu a_\nu\Big\}
\eqno(3.33)
$$
with $a_\mu(x)$ defined by (3.27) and (3.23)
for $\mu=0,1$. Notice that the integrand of (3.33)
is a total derivative and thus $S_{\rm top}$ is
a pure topological term which simply represents the flux
associated with the vector field $a_\mu(x)$.

We summarize the results obtained so far
by writing the full continuum action $S_{tJ}^{J=0}$, namely
$$
S_{tJ}^{J=0} = S_{\rm top} + \int\!d^2x~\Big\{
{\bar \psi}\big(\,{\slash\!\!\!\partial}
+{\rm i}\,{\slash \!\!\!a}\big)
\psi
+\Delta_\alpha^*\,\partial_0z_\alpha -
\partial_0z_\alpha^*\,\Delta_\alpha\Big\}\ \ .
\eqno(3.34)
$$

To complete our analysis now we switch on the spin-exchange
interaction ({\it i.e.} the $J$ term (2.35)). We assume for
simplicity that $J$ be small as compared to $t$.
(This is precisely the case of interest for the
phenomenological applications of the $t$-$J$ model.)
As we can see from (2.35), ${\cal H}_J$
is rather complicated: it contains quartic
terms both in the holon and in the spinon fields. Thus, to render
it tractable, we must make some approximations. To this
aim, let us first observe that the presence of a non zero
spin-exchange interaction certainly
modifies the hopping of the holons
with corrections of order $J$ to the $t$ term
considered so far; but since $J<\!<t$,
this effect is negligible.
Another consequence of a non zero $J$ term is
the appearance of a nearest-neighbor coupling between
spinons which may induce spin exchanges in the chain.
As is clear from (2.35),
if in two neighboring sites $a$ and $b$ no holon is present, then
the corresponding spinons interact and the spins in $a$ and $b$ can be
exchanged. But, if $a$ or $b$ or both are occupied by a holon, then
the $J$ term vanishes. This means that the spinon dynamics is heavily
influenced by the presence or the absence of the holons, and since
these are mobile, two neighboring spinons
may interact at some times but not at others. However, we
must remember that
$t>\!>J$, which means that
the holon motion is very quick as compared to the spinon
dynamics. Therefore, to describe the latter
it is reasonable to average over all
possible holon configurations and replace
${\cal H}_J$ with an effective spinon interaction,
in which the coupling constant is reduced with respect to
$J$. In particular, this amounts to replace the holon factors of
${\cal H}_J$ with the corresponding average values computed with the
tight-binding Hamiltonian (3.3) that is independent of the
spinons. More precisely inside ${\cal H}_J$ we perform
the following substitution
$$
\Big(1-
h^*(a)\,h(a)\Big)\Big(1-
h^*(b)\,h(b)\Big)\longrightarrow\left\langle
\Big(1-
h^\dagger(a)\,h(a)\Big)\Big(1-
h^\dagger(b)\,h(b)\Big)\right\rangle\ \ .
\eqno(3.35)
$$
Using Wick's theorem and (3.5) and (3.9), the right hand side
of (3.35) becomes
$$
1-2\,\delta+\delta^2 - {\sin^2(\pi\,\delta)\over \pi^2}\ \ .
$$
Thus, within this approximation we can replace ${\cal H}_{J}$ with
$$
{\tilde {\cal H}_J} = -{{\tilde J}\over 2}
\sum_{<a,b>} s^*_\alpha(a)\,s_\alpha(b)\,s^*_\beta(b)
\,s_\beta(a)\ \ ,
\eqno(3.36)
$$
where the renormalized spin-exchange coupling constant
${\tilde J}$ is defined by
$$
{\tilde J} = J\,(1-\delta)^2\left(1-{{\sin^2(\pi\,\delta)}\over {\pi^2\,
(1-\delta)^2}}\right)\ \ .
\eqno(3.37)
$$
Notice that ${\tilde J}$ is smaller than $J$ (as expected),
and, like $J$, is positive.
We remark that our approximation is different from the one that leads
to the squeezed spin chain considered in [14,29], even though there
are some similarities in the final formulas. In fact, in our case
the effective Hamiltonian ${\tilde {\cal H}_J}$ is still defined
on the original chain and not on a squeezed one like in [14,29]. This
is because the spinons in the ${\rm CP}^1$ representation
are defined everywhere, even where holons are present. However,
the differences between our approach and the squeezed spin chain
approximation disappear in the continuum limit.

To complete our analysis, we now
insert into (3.36) the decomposition (3.16) and keep all the terms
up to order $\ell^2$. (The square root factors of (3.16) are
now important since they yield contributions
of order $\ell^2$.) Upon expanding the products in (3.36),
we obviously produce terms quadratic
in $\Delta$, terms linear in $\Delta$
and terms
without $\Delta$. However, using (3.18), (3.19) and
(3.20), after some elementary algebra we may show that all the terms
linear in $\Delta$ exactly cancel each other, leaving us with the
following expression
$$
\eqalign{
{\tilde {\cal H}_J}=&-{{\tilde J\,\ell}\over 2}\Bigg\{
2\ell\sum_b\Big[{1\over 2}\partial_1^2 z_\alpha^*(b)\,z_\alpha(b) +
\partial_1 z_\alpha^*(b)\,z_\alpha(b)\, z_\beta^*(b)\,\partial_1 z_\beta(b)
+{1\over 2} z_\alpha^*(b)\,\partial_1^2 z_\alpha(b)\cr
&~~~~~~~~~~~~~~~~~
+4\,\Delta_\alpha^*(b)\,A^{\alpha\beta}(z,z^*)\,\Delta_\beta(b)
\Big]\Bigg\} -{\tilde J}\,N\ \ ,}
\eqno(3.38)
$$
where the matrix $A$ is explicitly given by
$$
A(z,z^*) =\left(\matrix{
 -z_2^*\,z_2 ~&~z_2^*\,z_1\cr
 {}&{}\cr
 z_1^*\,z_2 ~&~-z_1^*\,z_1\cr}\right)\ \ .
\eqno(3.39)
$$
It is amusing to observe that
$$
{1\over 2}\partial_1^2 z_\alpha^*\,z_\alpha +
\partial_1 z_\alpha^*\,z_\alpha\, z_\beta^*\,\partial_1 z_\beta
+{1\over 2} z_\alpha^*\,\partial_1^2 z_\alpha
= -\left|\big(\partial_1+{\rm i}\,a_1\big) \,{\bf z}\right|^2\ \ ,
\eqno(3.40)
$$
where in the right hand side we have introduced
a standard vectorial notation for the two-component vector
$$
{\bf z} =\left(\matrix{
z_1\cr
z_2 \cr}\right)\ \ .
\eqno(3.41)
$$
Let us define for convenience the quantity
$$
c = {\tilde J}\,\ell
\eqno(3.42)
$$
which has the dimensions of a velocity.
As we will see later, $c$ can be interpreted as the
characteristic velocity of the spin-waves
produced by the fluctuating spinons.

Dropping the irrelevant additive constant $-{\tilde J}\,N$ from
(3.38) and taking the continuum limit, we find that
the action associated to
${\tilde {\cal H}_J}$ is
$$
{\tilde S}_J = {c\over 2} \int \!d^2x
\left|\big(\partial_1+{\rm i}\,a_1\big) \,{\bf z}\right|^2
-2\,c\int \!d^2x ~
\Delta_\alpha^*\,A^{\alpha\beta}\,\Delta_\beta\ \ .
\eqno(3.43)
$$

If we collect everything together, we can write the
effective partition function of the $t$-$J$ model at small
$J$ as follows
$$
{\cal Z}_{tJ} = {\cal N} \int \!~{\cal D}^2\!\psi ~
{\cal D}^2{\bf z} ~{\cal D}^2\!\Delta_1
{}~{\cal D}^2\!\Delta_2~{\cal D}\lambda
{}~~{\rm e}^{-S_{tJ}}\ \ ,
\eqno(3.44)
$$
where ${\cal N}$ is a normalization factor and
$$
\eqalign{
S_{tJ} =& S_{\rm top} + \int\! d^2x~\Big\{
{\bar \psi}\big(\,{\slash\!\!\!\partial}
+{\rm i}\,{\slash \!\!\!a}\big)
\psi+
{c\over 2}\,\left|\big(\partial_1+{\rm i}\,a_1\big)
\,{\bf z}\right|^2 +\lambda\,\big(|{\bf z}|^2-1\big)\Big\}
\cr
&-\int \!d^2x~\Big\{2\,c
\,\Delta_\alpha^*\,A^{\alpha\beta}\,\Delta_\beta -
\big(\Delta_\alpha^*\,\partial_0z_\alpha -
\partial_0z_\alpha^*\,\Delta_\alpha\big) \Big\}\ \ .
}
\eqno(3.45)
$$
We remark that since we have taken the continuum limit, the integration
over both the ``background'' ${\bf z}$ and the fluctuations
$\Delta$ in (3.44) does not over count the
bosonic degrees of freedom, just like it
happens in the background field formalism
of quantum field theory.
The $\psi$-term in $S_{tJ}$ represents
the low-energy effective
action for the charge degrees of freedom. On the contrary,
the spinon terms
contain both long and short-distance effects, the former
described by ${\bf z}$ and the latter  by
$\Delta$. To remove
the short-distance effects and obtain the
low-energy effective action also for the spin degrees
of freedom, we must integrate out the fluctuations
$\Delta$. This integration is feasible since (3.45) is
simply a quadratic form in $\Delta$; in fact what we have to compute is
$$
{\tilde I} = \int\!{\cal D}^2\!\Delta_1~{\cal D}^2\!\Delta_2 ~
\exp\Bigg[\int\!d^2x~\big(2\,c
\,\Delta_\alpha^*\,A^{\alpha\beta}\,\Delta_\beta -
\Delta_\alpha^*\,\partial_0z_\alpha +
\partial_0z_\alpha^*\,\Delta_\alpha\big)\Bigg]\ \ .
\eqno(3.46)
$$
It is worth pointing out that the determinant of the matrix $A$
is zero and its trace is $-1$, as we can immediately see from
(3.39). This means that its eigenvalues are 0 and $-1$. Due to the
overall positive sign in the exponent (3.46), the presence of a
negative eigenvalue is welcome, but the presence of a zero mode
is disturbing since it may render ${\tilde I}$ divergent.
However, a careful analysis shows that this is not a real problem
because the (divergent) integral over the
zero mode turns out to be a constant
that can be reabsorbed into the normalization ${\cal N}$ of
the partition function. To see this explicitly, let us
denote by
$$
\Lambda^0_\alpha = z_\alpha ~~~~,~~~~
\Lambda^-_\alpha= -
\varepsilon_{\alpha\beta}\,z_\beta^*
$$
the eigenvectors of $A$ corresponding respectively to eigenvalues 0 and $-1$.
Then, let us decompose $\Delta_\alpha$ along these eigenvectors according to
$$
\Delta_\alpha = \Lambda_0\,\Lambda^0_\alpha + \Lambda_-\,\Lambda^-_\alpha\ \ ,
$$
where
$$
\Lambda_0 = z^*_\alpha\,\Delta_\alpha~~~~,~~~~\Lambda_-=
\varepsilon_{\alpha\beta}\,z_\alpha\,\Delta_\beta\ \ .
\eqno(3.47)
$$
With straightforward algebra, we can rewrite the exponent of (3.46) as
$$
\eqalign{
\int\!d^2x&~\Big( -2\,c\,|\Lambda_-|^2
-\Lambda_-^*\,\varepsilon_{\alpha\beta}\,z_\alpha\,\partial_0z_\beta
+\varepsilon_{\alpha\beta}\,z_\alpha^*\,\partial_0z_\beta^*\,
\Lambda_-
\cr
&~~~~~-\Lambda_0^*
\,z_\alpha^*\,\partial_0z_\alpha
+\partial_0z_\alpha^*\,z_\alpha\,\Lambda_0 \Big)\ \ .}
\eqno(3.48)
$$
Now, let us observe that the constraint (3.19) implies that
$$
\Lambda_0 = - \Lambda_0^*\ \ ,
$$
and thus the second line of (3.48) identically vanishes. Therefore,
the zero mode $\Lambda_0$ completely decouples from the dynamical
fields and we find
$$
\eqalign{
{\tilde I} &= \Big(\!\int\!{\cal D}^2\!\Lambda_0\Big)\!\!
\int \!{\cal D}^2\!\Lambda_-
\exp\Bigg[\!\int\!d^2x\,\Big(\!-2\,c\,|\Lambda_-|^2
-\Lambda_-^*\,\varepsilon_{\alpha\beta}\,z_\alpha\,\partial_0z_\beta
+\varepsilon_{\alpha\beta}\,z_\alpha^*\,\partial_0z_\beta^*\,
\Lambda_-\Big)\Bigg]\cr
&={\rm const} \,\cdot\,
\exp\Bigg[-{1\over 2\,c}
\int\!d^2x\,\Big(\varepsilon_{\alpha\beta}\,
z_\alpha^*\,\partial_0z_\beta^*\big)\,\big(\varepsilon_{\alpha\beta}\,
z_\alpha\,\partial_0z_\beta\Big)\Bigg]\ \ .
}
\eqno(3.49)
$$
It is straightforward to show that
$$
\big(\varepsilon_{\alpha\beta}\,
z_\alpha^*\,\partial_0z_\beta^*\big)\,\big(\varepsilon_{\alpha\beta}\,
z_\alpha\,\partial_0z_\beta\big) =
\left|\big(\partial_0+{\rm i}\,a_0\big) \,{\bf z}\right|^2
$$
which is the natural completion of the terms we found before.

Following the standard procedure, let us introduce the
spin-wave stiffness
$$
\rho = {{\tilde J}\,\ell\over 4}
\eqno(3.50)
$$
which in one space dimension is simply related to the spin-wave
velocity (cf (3.42)). Then, if we collect everything together
we can write the effective partition function for the $t$-$J$ model
at small $J$ in the following suggestive way
$$
{\cal Z}_{tJ} = {\cal N}' \int\!{\cal D}^2\!\psi~
{\cal D}^2{\bf z}
{}~{\cal D}\lambda
{}~~{\rm e}^{-S^{\rm eff}_{tJ}}\ \ ,
\eqno(3.51)
$$
where ${\cal N}'$ is a new normalization factor and
$$
\eqalign{
S^{\rm eff}_{tJ} =& S_{\rm top} + \int\! d^2x~\Bigg\{
{\bar \psi}\big(\,{\slash\!\!\!\partial}
+{\rm i}\,{\slash \!\!\!a}\big)
\psi+\lambda\,\big(|{\bf z}|^2-1\big)\cr
&~~~+2\,\rho\,\Big[\left|\big(\partial_1+{\rm i}\,a_1\big)
\,{\bf z}\right|^2 +{1\over c^2}\left|\big(\partial_0+{\rm i}\,a_0\big)
\,{\bf z}\right|^2\Big]\Bigg\}\ \ .
}
\eqno(3.52)
$$
{}From this expression it is clear that $c$ indeed
represents the velocity of
the spin-waves described by the spinon field ${\bf z}$
as mentioned above.
Of course, we are free
to choose our units in such a way that $c=1$ and the
last part of the action becomes formally ``Lorentz''
invariant. Notice however that in doing this, the Fermi
velocity of the holons which is used in the ``/'' notation
becomes
$$
v_{\rm F} = {2\,t\over {\tilde J}}\,\sin (\pi\,\delta)
\eqno(3.53)
$$
that in general differs from one. Therefore, our statement
about the ``Lorentz'' invariance of (3.52) must be
suitably interpreted. Of course this is not unexpected, because
in our problem there are two different characteristic fundamental
velocities: one for the charge degrees of freedom and one for
the spin degrees of freedom.

The action (3.52) is manifestly invariant under
the local gauge transformation (3.30) supplemented by
$$
z_\alpha(x) \longrightarrow z'_\alpha(x)={\rm e}^{-{\rm i}
\Lambda(x)}~z_\alpha(x)\ \ .
\eqno(3.54)
$$
Of course, (3.54) is compatible with (3.30), (3.27) and (3.23)
as one can immediately verify.
This local gauge invariance is not at all unexpected. Indeed, it has
its roots in the fact that the original factorization
of the electron operators in (2.6) leaves the freedom of
choosing arbitrarily the phase of holons and spinons in any point
of the lattice. This freedom then translates into the gauge
transformations (3.54) and (3.30) of the effective continuum
field theory.

It is interesting to remark that (3.52) is nothing but the action of
a ${\rm CP}^1$ model [15,16,17] with a coupling constant
proportional to $1/\rho$ and minimally coupled to
a fermionic field. Therefore, we can formulate our results
by saying that the effective dynamics of
the one-dimensional $t$-$J$ model at small $J$
in the ${\rm CP}^1$ representation
is described by a ${\rm CP}^1$ model for the spin degrees of freedom
minimally coupled to a massless
fermionic field representing the low-energy
charge excitations.

\vskip 2cm
\centerline{\bf 4. Conclusions and Outlook}
\vskip 0.7cm

In the previous section we managed to represent the
effective dynamics of the one-dimensional $t$-$J$ model
at small $J$
by means of a continuum field
theory with an explicit Abelian gauge invariance.
However, as we can see from
(3.27) and (3.23),
the vector potential $a_\mu$ appearing in the action
is not an independent
field as is customary in gauge theories. On the contrary,
it is a functional entirely determined by other fields, {\it viz.}
by the slowly varying spinons.
A similar situation occured also in the original formulation of
the ${\rm CP}^1$ model and its large-$N$ generalizations [16];
but there, with a suitable reinterpretation,
it was possible to introduce the gauge potential as an independent
degree of freedom, and then study its dynamics, for example using
a large-$N$ expansion.
We recall that within the framework of quantum field theory
the ${\rm CP}^1$ model and its
generalizations attracted a lot of attention
as interesting examples
of non trivial renormalizable
gauge theories exhibiting confinement,
dimensional
transmutation and topological effects
(for reviews see {\it e.g.} [17]).
It would be very interesting to explore these issues and study
the role of these
models also in the context of the strongly
correlated electron
systems; to this aim, a good starting point is certainly represented by
the action (3.52) that we derived directly from the $t$-$J$ model.

To make the connection with [16,17]
more explicit, let us fix $c=1$ in (3.52) and
perform a Wick rotation on $x^0$ to transform the two-dimensional
Euclidean space considered so far into a two-dimensional Minkowski space-time;
the resulting Lagrangian is then
$$
\eqalign{
L =&{1\over 2g^2}
\big(\partial^\mu-{\rm i}\,a^\mu\big){\bf z}^*\!\!\cdot\!
\big(\partial_\mu+{\rm i}\,a_\mu\big){\bf z}
+\lambda\,\big({\bf z}^*\!\!\cdot\!{\bf z}-1\big) + L_{\rm top}\cr
&+{\bar \psi}\big({\rm i}{\slash\!\!\!\partial}
-{\slash \!\!\!a}\big)\psi \ \ . }
\eqno(4.1)
$$
Here we have introduced a dimensionless
coupling constant $g^2$ for the ${\rm CP}^1$ term;
because of our choice of units, $g^2$ is actually fixed to be one
(cf (3.50)),
but it could be more convenient to think of it
as a parameter. Furthermore, in (4.1) the space-time
indices are contracted with the Minkowski metric $\eta_{11}=-\eta_{00}=1$,
and the gamma matrices used in the ``/'' symbol are
${\tilde \gamma}_0= {\rm i}\,\sigma^1$ and ${\tilde
\gamma}_1=v_{\rm F}\,\sigma^2$ with $v_{\rm F}$
given by (3.53). (For ease of notation we have incorporated
the Fermi velocity into the definition of ${\tilde
\gamma}_1$, and thus these gamma matrices satisfy a modified
Clifford algebra.) Finally, $L_{\rm top}$ is the topological
Lagrangian which is given by the integrand
of (3.33) also in a Minkowski space-time.

If, for a moment, we consider $a_\mu$ as an independent
field and compute its equation of motion from the first
line of the Lagrangian
(4.1) denoted by $L_z$, we get
$$
{{\delta L_z}\over {\delta a_\mu}}=0 ~~\Longrightarrow~~
a_\mu = {\rm i}\,{\bf z}^*\!\!\cdot\!\partial_\mu {\bf z} \ \ ,
\eqno(4.2)
$$
{\it i.e.} precisely the (Wick rotated)
definition given in (3.27) and
(3.23). Notice that $L_{\rm top}$,
being a topological term,
does not give any contribution to this field equation.
The result in (4.2) is quite interesting because it
shows that the wanted relation between the gauge field
$a_\mu$ and the bosonic vector ${\bf z}$ may be dynamically
determined since it could be viewed as an
equation of motion. However, (4.2) is not the true
field equation of $a_\mu$ determined by
the whole Lagrangian $L$, since also
the fermionic terms in the second line of (4.1) give
a contribution to it because the
vector potential $a_\mu$ couples to $\psi$. This
problem is easily cured if, instead of (4.1),
we consider the following Lagrangian
$$
\eqalign{
L' =&{1\over 2g^2}
\big(\partial^\mu-{\rm i}\,A^\mu\big){\bf z}^*\!\!\cdot\!
\big(\partial_\mu+{\rm i}\,A_\mu\big){\bf z}
+\lambda\,\big({\bf z}^*\!\!\cdot\!{\bf z}-1\big) + L_{\rm top}\cr
&+{\bar \psi}\big({\rm i}{\slash\!\!\!\partial}
-{\slash \!\!\!\!A}\big)\psi +{g^2\over 2}\,{\bar \psi}\,{\tilde\gamma}^\mu
\,\psi~{\bar \psi}\,{\tilde\gamma}_\mu\,\psi \ \ ,}
\eqno(4.3)
$$
where the vector field $A_\mu$ is
completely independent of the other fields, and
a new four-fermion interaction has been introduced.
Notice that $A_\mu$ appears in $L'$ without a kinetic term, and thus it
can be effectively eliminated through its equation of motion,
which is given by
$$
{{\delta L'}\over {\delta A_\mu}}=0 ~~\Longrightarrow~~
A_\mu = {\rm i}\,{\bf z}^*\!\!\cdot\!\partial_\mu {\bf z} +
g^2\,{\bar \psi}\,{\tilde\gamma}_\mu\,\psi = a_\mu + g^2\,{\bar
\psi}\,{\tilde\gamma}_\mu\,\psi \ \ .
\eqno(4.4)
$$
If we substitute this into $L'$, we exactly
reproduce the original
Lagrangian (4.1); in particular, the
four-fermion interaction of (4.3) has been designed
to cancel the fermionic terms arising when (4.4) is inserted
into $L'$.
Thus, we can say that (4.3) is equivalent to (4.1),
but it is more useful than the latter since
the gauge field $A_\mu$ in (4.3) is truly an
independent degree of freedom, whilst $a_\mu$
in (4.1) is
functionally detemined by the bosonic vector ${\bf z}$.
We point out that the four-fermion interaction of $L'$
is renormalizable in two dimensions
being marginal (indeed the coupling
constant in front of it is dimensionless).
It is one of the standard
fermionic interactions that are usually considered
in this context, the other being
$\big({\bar \psi}\,\psi\big)^2$ and
$\big({\bar \psi}\,\gamma_5\,\psi\big)^2$ where $\gamma_5=\sigma^3$.

The general structure of our final result is similar
to that of previous investigations on
the dynamics of holes in doped antiferromagnets; however a more
detailed comparison, in particular with [18,20], shows
important and significant differences.
First of all, contrarily
to [18,20], we took as our starting
point the $t$-$J$ model in the
${\rm CP}^1$ representation, and deduced
everything from it. The coupling between holons and spinons
is then determined by the structure of the Hamiltonian of
$t$-$J$ model and is not introduced by hand. In the low-energy limit,
our model turns out to be
described by the Lagrangian $L'$ (or better by its Euclidean
version) which, though similar, is different from that of [18,20]
in many respects.
Indeed, we have only one fermionic field describing the charge degrees
of freedom instead
of two, and moreover we have also a four-fermion interaction that
is not present in [18].
The reason for having one holon field instead of two is because
in the ${\rm CP}^1$ representation it is possible to
realize a nearest neighbor hopping of holons
in the presence of a ferromagnetic arrangement
of fictitious spinons, as we discussed at length in section
2. On the contrary, if one wants to describe
the charge motion in doped antiferromagnets without using
the ${\rm CP}^1$ representation, it becomes necessary to
introduce by hand a next-to-nearest neighbor holon
hopping term as it was done in [18,20], and consequently
to define two different types of holons each one moving
only within one sub-lattice.
However, a next-to-nearest neighbor hopping term is
vanishing from the point of view of the $t$-$J$ model
and hence it would be unnatural for us.

We conclude this paper by pointing out that if we generalize
the doublet (3.41) to an $N$-component vector
$$
{\bf z} =\left(\matrix{
z_1\cr
\vdots\cr
z_N \cr}\right)
\eqno(4.5)
$$
with ${\bf z}^*\!\!\cdot\!{\bf z}=1$, the action (4.3) becomes
that of a ${\rm CP}^{N-1}$ model coupled to one fermionic field. Since
${\bf z}$ appears quadratically in this action, it can be integrated out
to yield an effective action for the gauge field $A_\mu$ and the
fermionic field $\psi$.
This may be studied non-perturbatively with a large-$N$ expansion [35];
in particular one may examine if the factorization of the electrons
into holons and spinons is really
consistent and survives loop corrections; furthermore,
one may analyze the role of the topological term of the
action in connection with Haldane's conjecture [11,34], and
determine the structure of the renormalization group flow
and the nature of
the long-distance fixed points, and hopefully also shed some
light on the dynamical mechanisms of the charge-spin separation
from a field theory point of view.
We leave these issues as well as a more complete analysis of
the system described by (4.3) to future investigations.

\vskip 2cm
\centerline{\bf Acknowledgements}
\vskip 0.6cm
\noindent
I would like to thank Stefano Sciuto for many inspiring discussions,
and for a careful and critical reading of the manuscript.

\vfill
\eject

\centerline{\bf References}
\vskip 0.7cm
\item{[1]}J.G. Bednorz and K.A. M\"uller, {\it Z. Phys.} {\bf B64} (1986) 189;
\medskip
\item{[2]} For a review see for example A.P.
Balachandran, E. Ercolessi, G. Morandi and A.M.
Srivastava, {\it Hubbard Model and Anyon
Superconductivity} (World Scientific Publishing
Co., Singapore 1990) and references therein;
\medskip
\item{[3]} F.C. Zhang and T.M. Rice, {\it Phys.
Rev.} {\bf B37} (1988) 3754;
\medskip
\item{[4]} For a review see for example I. Affleck, in {\it
Fields, Strings and Critical Phenomena},
edited by E. Br\'ezin and J. Zinn-Justin
(North Holland, Amsterdam, The Netherlands 1990);
\medskip
\item{[5]} E.H. Lieb and F.Y. Wu, {\it Phys.
Rev. Lett.} {\bf 20} (1968) 1445;
\medskip
\item{[6]} P. Schlottmann, {\it Phys. Rev.}
{\bf B36} (1987) 5177; P.A. Bares, G.
Blatter and M. Ogata, {\it Phys. Rev.}
{\bf B44} (1991) 130; F.H.L. Essler and V.E. Korepin, {\it Phys. Rev.}
{\bf B46} (1992) 9147;
\medskip
\item{[7]}M. Ogata and H. Shiba, {\it Phys. Rev.}
{\bf B41} (1990) 2326; H. Shiba and M. Ogata, {\it Progr. Theor. Phys. Suppl.}
{\bf 108} (1992) 265;
\medskip
\item{[8]} P.W. Anderson, {\it Science}
{\bf 235} (1987) 1196; P.W. Anderson, in {\it Frontiers and
Borderlines in Many-Particle Physics}, edited
by R.A. Broglia {\it et al.} (North-Holland,
Amsterdam, The Netherlands 1988);
\medskip
\item{[9]} G. Baskaran, Z. Zou and P.W.
Anderson, {\it Solid State  Comm.}
{\bf 63} (1987) 973; Z. Zou and P.W.
Anderson, {\it Phys. Rev.} {\bf B37} (1988) 5594;
\medskip
\item{[10]}S.J. Barns, {\it J. Phys.} {\bf F6}
(1976) 1375; {\it J. Phys.} {\bf F7} (1977) 2637;
\medskip
\item{[11]}F.D.M. Haldane, {\it Phys. Lett.} {\bf 93A} (1983) 464;
{\it Phys. Rev. Lett.} {\bf 50} (1983) 1153;
\medskip
\item{[12]}I. Affleck, {\it Nucl. Phys.} {\bf B257} (1985) 397;
{\it Nucl. Phys.} {\bf B265} (1985) 409;
\medskip
\item{[13]}P.B. Wiegmann, {\it Phys. Rev. Lett.} {\bf 60}
(1988) 821; {\it Physica} {\bf C153-155} (1988) 103;
L.B. Ioffe and P.B. Wiegmann, {\it Phys. Rev. Lett.} {\bf 65}
(1990) 653;
\medskip
\item{[14]}Z.Y. Weng, D.N. Sheng, C.S. Ting and Z.B. Su, {\it Phys. Rev.}
{\bf B45} (1992) 7850; {\it Phys. Rev. Lett.} {\bf 67} (1991) 3318;
\medskip
\item{[15]}H. Eichenherr, {\it Nucl. Phys.} {\bf B146} (1978) 215;
V. Golo and A. Perelomov, {\it Phys. Lett.} {\bf 79B} (1978) 112;
\medskip
\item{[16]}A. D' Adda, P. Di Vecchia and M. L\"uscher,
{\it Nucl. Phys.} {\bf B146} (1978) 63; {\it Nucl. Phys.}
{\bf B152} (1979) 125; E. Witten,  {\it Nucl. Phys.}
{\bf B149} (1979) 285;
\medskip
\item{[17]}S. Coleman, in {\it Aspects of Symmetry} (Cambridge
University Press, Cambridge, England 1985);
A.M. Polyakov, in {\it Gauge, Fields and Strings}
(Harwood Academic Publishers,
Chur, Switzerland 1987);
\medskip
\item{[18]}R. Shankar, {\it Phys. Rev. Lett.} {\bf 63}
(1989) 203; {\it Nucl. Phys.} {\bf B330} (1990) 433;
\medskip
\item{[19]}P.A. Lee, {\it Phys. Rev. Lett.} {\bf 63} (1989) 680;
\medskip
\item{[20]}X.G. Wen and A. Zee, {\it Nucl. Phys.} {\bf B326} (1989) 619;
\medskip
\item{[21]}N. Dorey and N.E. Mavromatos, {\it Phys. Rev.} {\bf B44} (1991)
5286;
\medskip
\item{[22]}G. Shirane {\it et al.}, {\it  Phys. Rev. Lett.} {\bf 59}
(1987) 1613;
\medskip
\item{[23]} M. Gutzwiller, {\it Phys. Rev.} {\bf A137}
(1965) 1726;
\medskip
\item{[24]}F.D.M. Haldane, {\it Phys. Rev. Lett.} {\bf 45}
(1980) 1358; {\it J. Phys.} {\bf C14} (1981) 2585; P.W. Anderson, {\it
 Phys. Rev. Lett.} {\bf 64} (1990) 1839;
\medskip
\item{[25]}H. Yokoyama and M. Ogata,
{\it Phys. Rev. Lett.} {\bf 67} (1991) 3610;
Y.R. Wang, {\it Phys. Rev.} {\bf B43} (1991)
3786; Y.R. Wang, J. Wu and M. Franz, {\it Phys. Rev.}
{\bf B47} (1993) 12140;
\medskip
\item{[26]}W.O. Putikka {\it et al.},
Z\"urich ETH preprint ETH-TH/93-33
(September 1993) cond-mat/9309031;
\medskip
\item{[27]}For reviews see for instance F. Wilczek,
in {\it Fractional Statistics
and Anyon Superconductivity} edited by F. Wilczek
(World Scientific Publishing Co., Singapore 1990);
A. Lerda, {\it Anyons: Quantum Mechanics of Particles
with Fractional Statistics} (Springer-Verlag,
Berlin, Germany 1992);
\medskip
\item{[28]} A. Lerda and S. Sciuto,
{\it Nucl. Phys.} {\bf B410} (1993) 577;
\medskip
\item{[29]} S. Feng, Z.B. Su and L. Yu, {\it Phys. Rev.} {\bf B49} (1994) 2368;
G.-M. Zhang, S. Feng and L. Yu, ICTP preprint (January 1994) cond-mat/9401070;
\medskip
\item{[30]}N. Read and S. Sachdev,
{\it Nucl. Phys.} {\bf B316} (1989) 609;
\medskip
\item{[31]}Y. Nagaoka, {\it Phys. Rev.} {\bf 147} (1966) 392;
\medskip
\item{[32]}See for example E. Fradkin, {\it Field Theories of
Condensed Matter Systems} (Addison - Wesley, Reading,
MA, USA 1991) and references therein;
\medskip
\item{[33]}D. Schmeltzer, {\it Phys. Rev.}
{\bf B42} (1990) 1074; {\it Phys. Rev.} {\bf B43} (1991) 8650;
\medskip
\item{[34]}R. Shankar and N. Read, {\it Nucl. Phys.} {\bf B330} (1990) 457;
\medskip
\item{[35]}I. Affleck and B. Marston, {\it Phys. Rev.} {\bf B37} (1988) 3774;
{\it Phys. Rev.} {\bf B39} (1989) 11538; D. Arovas and A. Auerbach,
{\it Phys. Rev.} {\bf B38} (1988) 316.
\vfill \eject

\centerline{\bf Figure Captions}
\vskip 1cm
\item{Fig. 1a}An electronic configuration in which the even site $a$ is
occupied by an electron of spin up and the nearest
neighbor odd site $b$ is empty.
\bigskip
\item{Fig. 1b}After hopping, the spin up electron has moved to $b$ and the hole
to $a$.
\vfill
\eject
\nopagenumbers
\hskip 9cm \vbox{\hbox{DFT-US 1/94}
\hbox{March 1994}}
\vskip 1.5cm
\centerline{{\bf A FIELD THEORY APPROACH TO THE $t$-$J$ MODEL}}
\centerline{{\bf AND THE SPIN-CHARGE SEPARATION }~\footnote{$^*$}{
Work supported in part by Ministero dell' Universit\'a e
della Ricerca
Scientifica e Tecnologica.}}
\vskip 1cm
\centerline{{\bf Alberto Lerda}~\footnote{$^a$}{e-mail address:
vaxsa::lerda or lerda@toux40.to.infn.it}}
\vskip 0.4cm
\centerline{\sl Dipartimento di Fisica Teorica, Universit\'a di Salerno}
\centerline{\sl I-84100 Salerno, Italy}
\vskip 0.2cm
\centerline{\sl and I.N.F.N.
Sezione di Napoli}
\vskip 2.5cm
\centerline{{\bf Abstract}}
\vskip 0.7cm
\noindent
We analyze the $t$-$J$ model using the ${\rm CP}^1$ representation
for the slave operators (holons and spinons) which
is particularly suited to study the phenomenon of the spin-charge
separation in strongly correlated electron systems.
In particular, we show that for the one-dimensional $t$-$J$
model below half-filling the low energy effective dynamics
of the spin and charge degrees of freedom is represented
in the continuum limit by a ${\rm CP}^1$ model with
a topological term, minimally coupled to a massless Dirac
field with a four-fermion interaction.
The bosonic term of this action describes the spin waves
produced by the spinons, while the fermionic term represents
the low energy charge excitations. This theory exhibits
explicitly a local abelian gauge invariance.
\bye
\end